# Hydro-mechanical simulation and analysis of induced seismicity for a hydraulic stimulation test at the Reykjanes geothermal field, Iceland

**Eirik Keilegavlen[1], Laure Duboeuf[2], Anna Maria Dichiarante[2], Sæunn Halldórsdóttir[1], Ivar Stefansson[1], Marcel Naumann[3], Egill Árni Guðnason[4], Kristján Ágústsson[4], Guðjón Helgi Eggertsson[5], Volker Oye[2], Inga Berre[1]**

[1]Department of Mathematics, University of Bergen, Norway

[2]Department of Applied Seismology, NORSAR, Norway

[3]Equinor ASA, Norway

[4]ÍSOR Iceland Geosurvey, Iceland

[5]HS Orka, Iceland

Corresponding author: Inga Berre, inga.berre@uib.no

**Key Points:**

- Framework for integrating different data types in development of a 3D faulted geothermal reservoir model

- Novel application of simulation tool for coupled hydro-mechanical processes in faulted geothermal reservoirs applied to injection-induced fault reactivation at Reykjanes, Iceland

- New insights through combined induced seismicity analysis and physics-based simulation of injection-induced fault reactivation

**Abstract**

The hydraulic stimulation of the well RN-34 at the Reykjanes geothermal field in Iceland caused increased seismic activity near the well. Here, we use this as a case study for investigation on how seismic analysis can be combined with physics-based simulation studies to further understand injection-induced fault reactivation. The work presents new analysis of the seismic data combined with application of a recent simulation software for modeling of coupled hydro-mechanical processes and fault deformation caused by fluid injection. The simulation model incorporates an explicit model of the fault network based on geological characterization combined with insights from seismic analysis. The 3D faulted reservoir model is then calibrated based on injection data. Despite limited data, the work shows how seismic interpretations can be used in developing simulation models and, reciprocally, how the modeling can add to the seismic interpretations in analysis of dynamics.

**1 Introduction**

Most of the Earth's accessible geothermal energy is stored in hard, competent rock. In such rock types, fractures and faults are the main conduits for fluid flow, which is essential for production of geothermal fluid to the surface. To enhance permeability in such formations, fluids at elevated pressures can be injected to cause slip and dilation of existing fractures or faults. This stimulation mechanism, called hydroshearing, has proved successful for permeability enhancement for several geothermal reservoirs (Chabora et al., 2012; Genter et al., 2010; Schindler et al., 2010; Zimmermann & Reinicke, 2010). This stimulated slip and dilation is often realized as microseismic events, or microearthquakes, emitting seismic waves. These waves of small elastic deformations can be recorded by local seismic networks; hence, the process of hydroshearing can be continuously detected, located, and analyzed in real time. In general, the stimulation aims to induce only small seismic events of magnitudes $M_w < 2$ (Ellsworth, 2013), but larger events have also been linked to hydraulic stimulation of fractured geothermal reservoirs. The most prominent are the 2017 $M_w$ 5.4 Pohang earthquake in South-Korea (Ellsworth et al., 2019; Grigoli et al., 2018; Kim et al., 2018) and the 2006 ML 3.4 earthquake related to the Basel EGS project (Bachmann et al., 2011; Deichmann & Giardini, 2009). In Iceland, the most prominent induced seismicity has been two magnitude 4 events that were observed related to geothermal wastewater reinjection in Húsmúli at the Hellisheidi geothermal area, with flow rates reaching 500 kg/s (Juncu et al., 2020).

Hydroshearing of fractures occurs in an interplay between coupled hydraulic, thermal, mechanical, and chemical reservoir processes and the fractured structure of the formation. To design hydraulic stimulation operations while mitigating induced seismicity, understanding of these coupled, nonlinear dynamics is crucial. In this, physics-based numerical models can provide valuable insights: either as a tool to forecast outcomes of a stimulation or to complement dynamic data in understanding the governing mechanisms and structural features at depth.

Several physics-based numerical modeling studies have considered how an increase in pore pressure reduces effective stress on preexisting fractures or faults, thus causing slip (Bruel, 2007; Gischig & Wiemer, 2013; Goertz-Allmann et al., 2011; Hakimhashemi et al., 2014; Kohl & Mégel, 2007; Rothert & Shapiro, 2003; Shapiro, 2015). To incorporate the important effect of stress redistribution due to hydroshearing is challenging; even the more advanced numerical models are typically based on strong simplifications of the physics governing flow, mechanics, or coupled hydromechanics related to the fractures and/or the domain surrounding them (McClure & Horne, 2011; Norbeck et al., 2016; Ucar et al., 2017; 2018a). Recently, however, numerical modeling tools which consider networks of fractures or faults in 3D domains have been developed that consistently account for fully coupled hydro-mechanical processes as well as fracture-contact mechanics (Berge et al., 2020; Gallyamov et al., 2018; Garipov et al., 2016; Garipov & Hui, 2019; Keilegavlen et al., 2019). In this paper, we investigate how simulations based on such numerical models can complement dynamic data in an investigation of hydroshearing, considering a specific case study for a stimulation test at the Reykjanes geothermal field in SW Iceland on 29 March 2015.

The large-scale exploitation of the Reykjanes field for geothermal energy started when Hitaveita Suðurnesja (Reykjanes District Heating), now HS Orka, acquired the development concession rights for the geothermal field and drilled its first well for electrical generation in 1998. In 2006, production started at the 100 MWe Reykjanes power plant. As of 2019, a total of 37 wells have been drilled in Reykjanes for exploration, production, and re-injection. The conceptual model of the geothermal system is described by Khodayar et al. (2018), Weisenberger et al. (2019), and Nielsson et al. (2020).

In 2014 and 2015, the wells RN-33 and RN-34 were drilled from a well pad northwest of Sýrfell, about 2 km northeast of the center of the main production area (Figure 1c). The wells were intended for re-injection of separated brine from the Reykjanes Power Plant. Well RN-33 is directionally drilled to the SW and connected to the production field through a NE-SW trending fissure zone. Well RN-34 is directionally drilled to the NW and results of tracer tests indicate that the well is not hydraulically connected to the production field. For RN-34 a fall-off test followed by 10 hours of cyclic stimulation was conducted on 29 March 2015 (see Supporting Information, Texts S3 and S4). Seismic events related to the injection were observed near the injection point.

During the period of interest, this seismicity at Reykjanes was recorded by both a permanent and a temporary seismic network (Figure 1b). The permanent network (PS) was run until 2018 by the Iceland Geosurvey (ISOR) on behalf of HS-Orka and was composed of eight short-period sensors mostly located above the geothermal reservoir on the SW part of the Reykjanes Peninsula (Weemstra et al., 2016). A temporary network (TS) with 20 broadband and 10 short–period sensors covering the entire Reykjanes Peninsula was installed as part of the European Project IMAGE and took recordings from March 2014 to August 2015 (Blanck et al., 2020; Jousset et al., 2016). The variety of sensor types and the short (0.8 km) and long (35 km) interstation distances increase network resolution and the capability of recording close and far events. Induced events have been observed at Reykjanes since the start of the geothermal activity (Blanck et al., 2020; Flovenz et al., 2015; Guðnason, 2014).

In this paper, we present an investigation of the hydraulic stimulation of RN-34 based on new analysis of the seismic data combined with an unprecedented simulation study of the reservoir dynamics. Based on all available data relevant to the study of the hydraulic stimulation test, we develop a novel hydro-mechanical model of the faulted reservoir to simulate the subsurface dynamics that occur as a response to hydraulic stimulation. For the simulation, we employ a simulator constructed for fully coupled flow, poroelasticity, and fracture deformation (Keilegavlen et al., 2019). The simulation model accounts for flow in both explicitly represented faults and the low-permeable surrounding porous medium, slip of faults based on a Coulomb friction law, and coupled poroelastic response of the porous medium to fluid pressure and fault slip. To our knowledge this represents the first application of a simulator constructed for fully coupled poroelasticity and fracture deformation to model stimulation of an actual geothermal reservoir.

The paper is structured as follows: In section 2, we present the regional context of the 29 March 2015 RN-34 stimulation test, including regional information for the stress state and a model of the dominating fault geometry near RN-34. Section 3 presents dynamic observations related to the RN-34 stimulation, including dynamic well data and analysis of induced seismicity. In section 4, we present the mathematical model and the simulation model, including parameter identification based on well data. Section 5 presents the numerical model and simulation results. A discussion of the combined results from seismic analysis and physics-based simulations is given in section 6, followed by a summary and concluding remarks in section 7.

## 2 From regional geological context to local fault model geometry

To model the coupled dynamics of a geothermal reservoir and analyze microseismic data, consideration of the target area's geological setting is important, including how the setting dictates today's local stress field conditions. In this section, we first introduce the regional geological setting: the larger structures and stresses that surround our target area. Then, we bring these into context with existing local-scale geological interpretations and recent seismological observations, which are integrated to understand the stresses that acted at a given time on certain faults involved in the local model.

The fault and fracture orientation reading convention used in this paper is strike (0°–360°)/dip (0°–90°) and rake (hanging-wall slip vector is measured on the plane of the fault) for recording fault planes/focal planes and their relative kinematics and trend (0°–360°)/plunge (0°–90°) for stress axes. Both conventions follow the right-hand rule, and both stereonets and focal mechanisms are projected on the lower hemisphere.

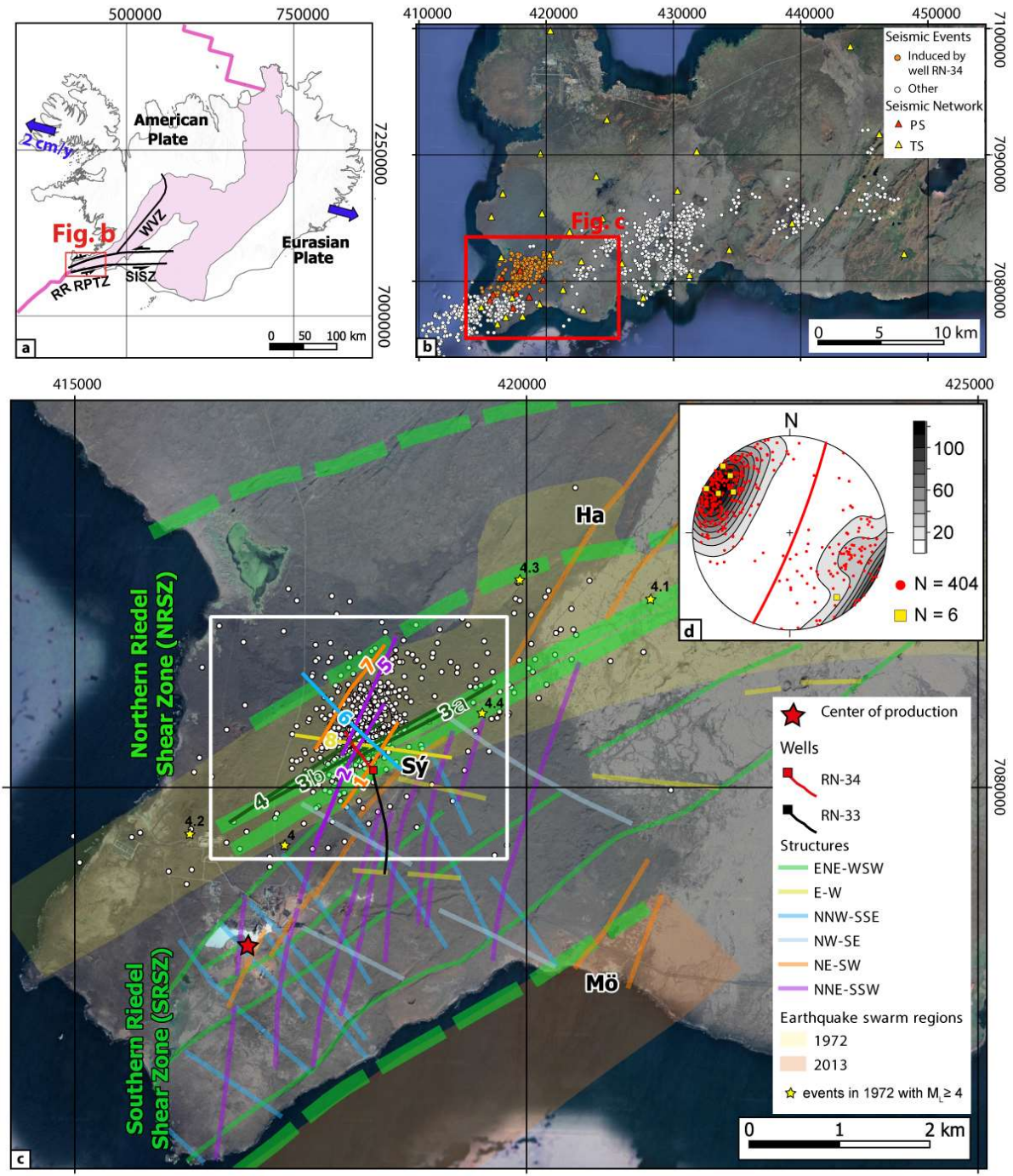

**Figure 1. (a)** Plate boundary across Iceland showing the study location (red box); the Reykjanes Peninsula Transtensional Zone (RPTZ) is located in the region where the Reykjanes Ridge (RR), South Iceland Seismic Zone (SISZ), and the Western Volcanic Zone (WVZ) meet. The plate boundary (line) and rocks of age less than 0.8 Ma (adapted from Khodayar et al., 2018) are shown in pink. **(b)** The Reykjanes Peninsula showing the two networks (yellow and red triangles, respectively) for the temporary seismic (TS) network and the permanent seismic (PS) network as well as the re-located seismic events (shown by white and orange dots). The events analyzed in this study are those shown by yellow dots. **(c)** Simplified structural map of the Reykjanes geothermal field adapted from Khodayar et al. (2018) showing boundaries of the two major Riedel shear zones: the Northern Riedel Shear Zone (NRSZ) and the Southern Riedel Shear Zone (SRSZ). Also shown are the outlines of the 1972 (yellow) and 2013 (orange) earthquake swarms as well as events from the 1972 swarm (Klein et al. 1977, Björnsson et al., 2020) with $M_{L \geq} 4$ (yellow stars). Structures are shown as transparent lines in the SRSZ and elsewhere and the initial considered fault model is shown using non-transparent lines in the studied area. For both, the color scheme is based on orientation. Induced seismic events are shown in white. The locations Sýrfell (Sý), Mölvík (Mö), and Haugur (Ha) are shown. The wellhead location of wells RN-34 and RN-33 is marked with a square and the well paths with lines. The approximate location of the center of production is indicated by a red star and the study area by a white rectangle. **(d)** Equal area stereonet plot and density contour of the fractures interpreted from well RN-34 by Árnadóttir et al. (unpublished report, Supporting Information, Text S1). Fractures are represented as pole to planes (red dots), and fractures with a mechanical aperture >19 mm as yellow squares (Supporting Information, Table S2). Mean orientation of fractures interpreted with high confidence is represented by a large red circle (Supporting Information, Table S1).

    2.1 Regional geological setting

Iceland is located at a complex mid-ocean ridge plate boundary between the Eurasian and American plates (Figure 1a), coinciding with a relatively large amount of hot, upwelling mantle material, which explains its volcanic activity. The Reykjanes geothermal field is located on the southwest tip of the Reykjanes Peninsula Transtensional Zone (RPTZ) in SW Iceland, where the Reykjanes Ridge (RR) comes onshore and the plate boundary changes direction. Rifting becomes oblique on the RPTZ and the rift segments split into a series of NE-SW trending eruptive fissures, which can be grouped into four en echelon volcanic fissure swarms (Sæmundsson, 1978). The fissure swarms, from east to west, Hengill, Brennisteinsfjöll, Krýsuvík and Reykjanes-Eldvörp-Svartsengi (for location see Fig. 1 in Keiding et al., 2009), consist of normal faults and tension fractures in addition to the eruptive fissures. They are intersected by a series of near vertical N-S trending right-lateral strike slip faults (Keiding et al., 2009). The regional extension direction of this oblique rift (extension is not orthogonal to plate boundary) is N101-103°E with an extension rate of 19–20 mm/yr along both active rift and transform segments (e.g., Keiding et al., 2009).

The patterns of natural seismicity are valuable in understanding the complex geological structures of the Reykjanes Peninsula (see, e.g., Björnsson et al., 2020, and Keiding et al., 2009). Based on GPS measurements taken during the 1993 and 1998 seismic swarm events, which showed almost exclusively strike-slip deformation, Clifton and Kattenhorn (2006) interpreted the geological structural complexity in Reykjanes, concluding that, to accommodate oblique spreading, episodes of tectono-magmatic activity (during extension) and episodes of left-lateral strike-slip motion alternate with different periodicity and in different structural blocks. However, fault plane solutions computed from two of many earthquake swarms on the Reykjanes Peninsula (in 1972 and 2013) showed that both normal, strike–slip, and oblique motions occurred during both earthquakes in a magmatic phase (Björnsson et al., 2020; Khodayar et al., 2018), supporting the idea that deformation might occur simultaneously on differently oriented structures. Furthermore, these two earthquake swarms helped to delineate two sinistral Riedel shear zones (ENE-striking boundary structures represented in green in Figure 1c). The largest seismic events in 1972 ($M_L > 4$) were located at the boundary (up to 2.6 km wide) between these Riedel shear zones (Figure 1c).

A comprehensive paper by Khodayar et al. (2018)—which integrated remote sensing, field geology, and seismicity—showed that both the NRSZ and the SRSZ, located within the Reykjanes geothermal field, are populated by a series of minor and differently oriented structures and that the SRSZ block is more intensely fractured than the NRSZ block. The authors categorized these structures in terms of their orientation and kinematics (Figure 1c): ENE-striking structures, which are subparallel to the NRSZ boundary, display sinistral sense of shear; N-striking to NNE-striking structures, which are mainly dextral strike-slip; NNE-striking to NE-striking structures, that bound for example two grabens at Haugur (Ha) and Mölvík (Mö), are mainly extensional faults and dikes intruding along NE-oriented fissures and faults; E-W structures, which, with significant uncertainty, are inferred to be dextral; and NW-striking to NNW-striking and WNW-striking structures, which are dextral (Figure 1c; Khodayar et al., 2018). Dextral N-striking to NNE-striking structures in the SRSZ are difficult to observe at surface; hence, they are mainly mapped from earthquakes (Keiding et al., 2009). These are interpreted to represent a conjugate fault system together with the ENE-striking structures (Khodayar et al., 2018) and are observed to cut across NE-striking volcanic fissures and normal faults. Both these volcanic fissures and normal faults accommodate extension while N-S faults accommodate the transform component (Sæmundsson et al., 2020).

### 2.2 Preliminary fault model

In this section, we extract a local fault model for our case study in which we incorporate measurements from wells and local geological studies. The study area is located at Sýrfell (Sý in Figure 1c), approximately 2 km NE of the center of production (Figure 1c), mainly in the NRSZ and partially across the boundary between the two shear zones (Figure 1c). Structural information on this region relies on fault traces interpreted by Khodayar et al. (2018) and well data and televiewer interpretation from well RN-34 (see Supporting Information, Text S1). Although a detailed outcrop study of the area is missing, movements along the faults (or kinematics) are believed to mirror the overall structural pattern of Reykjanes. Furthermore, televiewer images from well RN-34 provide valuable information on orientation, infill, (mechanical) aperture, and kinematics of fractures intercepted by the well. A total of 404 N-striking and NE-striking fractures were interpreted in an unpublished report by Árnadóttir et al. (unpublished report, Supporting Information, Text S1). The dominating fractures (interpreted with high confidence) are subvertical (83°) and strike NNE (022°) on average. An apparent mechanical aperture larger than 19 mm was measured for a series of NNE-striking to NE-striking fractures.

Due to the opening of the NNE-striking to NE-striking fractures and their vicinity to feed points, Árnadóttir et al. (unpublished report, see Supporting Information, Text S1) assume that four of these fractures act as fluid pathways.

In an unpublished report, Khodayar et al. (Supporting Information, Text S2) suggested five preliminary fault models consisting of nine structures (labelled 1 to 8 and 3b in the white rectangle in Figure 1c), representing the starting point of our local fault model for the case study. Two of these faults have known dips (75° to NW) from the outcrop study; they did not intersect the well and are far from the seismic cloud. These faults were therefore excluded from the fault model. In the models, the remaining faults are believed to have constant dip, as do all of the faults in each model (70°, 75°, or 90°). The NW-oriented fault trace (no. 6 in Figure 1c), together with a similarly oriented lineament to the east (not shown in Figure 1c), is, according to Khodayar et al. (2018), likely bounding the seismic cloud of the 2015 swarm (including events occurring on 12 December 2014). This structure is unfavorably oriented to slip since it is orthogonal to the $\sigma_1$ axis (as discussed in sections 2.3 and 2.4). We assume that it has limited impact on flow. Therefore, it has also been excluded from the fault model.

### 2.3 Revised fault model based on the analysis of induced seismicity

In this section we incorporate observations from seismicity clouds and focal mechanisms to generate the local fault model geometry. Using locations of induced and naturally occurring seismicity is a well-suited alternative for the identification of faults where a lack of clear reflectors and large impedance contrasts impair the value of conventional active seismic methods. One benefit of analyzing induced seismicity is that only active faults will be identified; active faults are likely to have relatively large aperture and permeability and are thus suitable for geothermal exploration. Since seismicity will also be induced during fluid injection, the respective locations of seismic events can be utilized to further constrain the geometry of the local fault model.

### 2.3.1 Brief overview of the methods used to interpret seismicity

Together, individual earthquakes and clouds of smaller seismic events contain two main pieces of information that can be used to construct and improve fault models: i) The relative location of a seismic cloud reveals the general fracture orientation (strike and dip), a.k.a. seismic lineation. Statistical analysis of the cloud (e.g., through the collapsing method [Fehler, 2000]) can further provide uncertainty estimates of the fracture orientation. ii) Focal mechanisms can be determined from larger individual events to constrain the orientation and kinematics of causative slipping fractures.

We employ the collapsing method to interpret fracture orientation from the cloud of seismic events. Introduced by Jones and Stewart (1997), this statistical method entails moving event locations iteratively within their relative uncertainty until the desired accuracy is obtained. Each event is represented by a point with a confidence ellipsoid. This ellipsoid will generally contain a cloud of other events, and an event is moved toward the centroid of the point cloud by a fraction of the event-centroid distance. By repeating this procedure, all the events are "collapsed" within their uncertainty ellipsoids, with the exception of very isolated events. Then, we apply a plane fitting method based on the computation of the covariance matrix of the collapsed events' point cloud to extract strike and dip and compare this geometry to the fault plane geometry already in the model.

To determine fault plane slipping during a seismic event, focal mechanisms are computed. These graphical representations provide two orthogonal fault plane solutions, of which only one is the correct, active or causative fault. Discriminating between the two solutions often relies on other information (e.g., distribution in space of the focal mechanisms, fault mapped at surface, etc.). Constraining focal mechanisms of small events can be more challenging when the amplitudes of first arrivals are small and the noise levels are high, as is often the case when investigating microseismic events with surface networks.

### 2.3.2 Interpretation of seismicity

Here, we focus on the seismic events occurring from 20 May 2015 to the end of the TS network recording, i.e., 13 August 2015 (the yellow region in Figure 2a). Automatic detection was applied on continuously recorded seismic data and led to ~6500 seismic events. About 3000 of these events were automatically picked based on pattern matching identification (Duboeuf, Oye, Berre, Keilegavlen, and Dando, 2019) and were quality controlled visually. The entire set of picked events was located using the Icelandic 1-D layer velocity model South Iceland Lowland (SIL, Bjarnason et al., 1993) and a differential evolution algorithm (Storn & Price, 1997; Wuestefeld et al., 2018).The location accuracy was increased using a Double-Difference relative location method (Waldhauser, 2000) (Figure 1b). A detailed analysis of seismic processing methods can be found in Duboeuf, Oye, Berre and Keilegavlen. (2019). Seismic events were grouped into several families based on waveform similarities and event locations. One group of 687 events was likely related to a fluid injection in injection well RN-34 (Figure 1b). The moment magnitudes ($M_w$) of these events range from about 0.8 to 3 and were determined by fitting a Brune model (Brune, 1970) to the observed seismic spectra. The resulting frequency-magnitude distribution follows the Gutenberg-Richter law, characterized by a high b-value (1.47), as often observed in fluid injection areas (Eaton et al., 2014; Shapiro et al., 2013).

Furthermore, seven time periods display a distinct increase in the seismic activity that surpasses the daily average (> 10 events/day); these are numbered 1 to 7 in Figure 2a. We refer to this spatially and temporarily limited increase in seismic activity as "bursts." Bursts occur within short time intervals (from a few hours to one week) and are concentrated in relatively small spatial regions, supporting the idea that they might be caused by slip on the same structure.

The collapsing method was applied to all the bursts (1 to 7 in Figure 2a), with the result indicating that six out of seven fitting planes are ENE-striking (ellipses in Figure 2b). Strike and dip of the planes are reported in Figure 2b, showing larger variability in the dip than the strike. This is likely attributable to greater uncertainty in the depth of these events due to the lack of sensors at depth, resulting in poorly constrained dip of the faults. Bursts with a larger number of events (numbers 1 and 4 with 95 and 313 events, respectively) suggest steep (81° to 88°) fault planes. The closest similarly oriented structure to these fitting planes is Fault 4 (Figure 1c). However, because of the proximity (approx. 200 m) of the two other, similarly oriented structures (Faults 3a and 3b, Figure 1c), we cannot rule out that the causative structure could also be one of those.

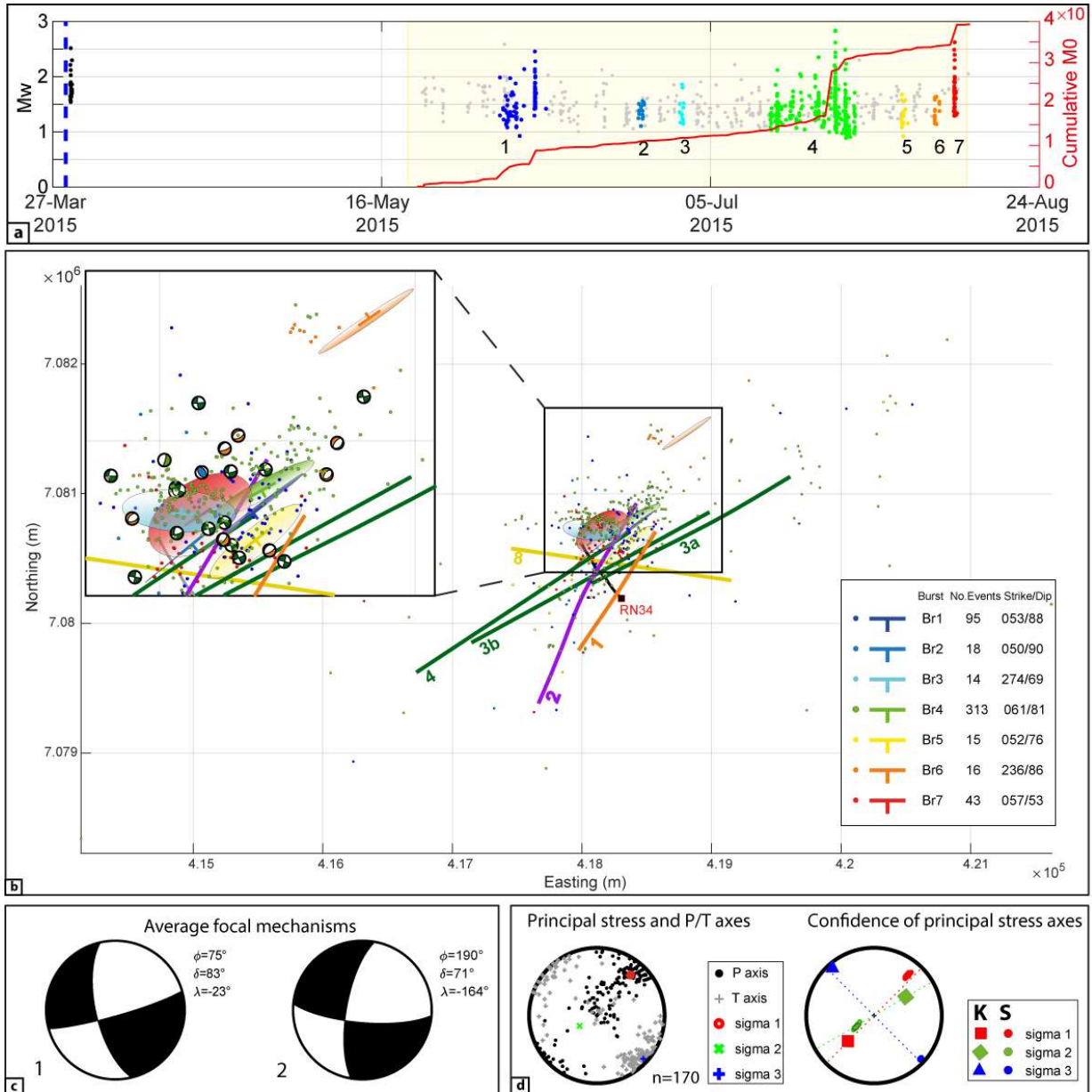

**Figure 2. (a)** Time vs. moment magnitude (Mw) plot showing post-stimulation seismicity at well RN-34. White regions represent periods where no data are available. Seven periods of increased seismicity for the period between 20 May and 13 August 2015 are shown as colored dots, numbered from 1 to 7. Cumulative seismic moment (M0) during the same period is also shown (red curve). **(b)** Collapsed bursts of seismicity (dots in the same color scheme as used in Figure 2a), fitted planes (ellipse and strike/dip symbol), and well-resolved focal mechanisms occurring in the same region. Focal mechanisms are colored by fault type: orange for normal, blue for reverse, and green for strike-slip. Well RN-34 is also shown (grey). **(c)** Average of 170 focal mechanisms showing strike–slip kinematics, with Φ denoting strike, δ dip, and λ rake. **(d)** Left: principal stress and P/T (Pressure or compression/Tension) axes; Right: confidence of principal stress axes inverted from 170 focal mechanisms (S) and stress orientation from Keiding et al. (2009). Note that P/T axes correspond to $\sigma_1$ (sigma 1) and $\sigma_3$ (sigma 3) axes.

170 focal mechanisms were computed from P-wave first motion polarities. The focal mechanism solutions indicate that the prevailing faulting types are strike-slip (52%) with near-vertical focal planes, normal (19%), and normal oblique (15%). Purely thrust (5%) and strike-slip with thrust component (2%) fault motions have also been identified. Finally, 7% of the determined focal mechanisms present a non-double-couple component, which might be expected in an injection area (Julian et al., 1998; Zhao et al., 2014). In the inset in Figure 2b, a selected number of focal mechanisms of high quality are superimposed on the collapsed planes. The quality of these focal mechanisms is normalized and based on the number of solutions, observations, and misfits. However, the locations of these focal mechanisms do not clearly align along one specific structure. The two most recurrent focal mechanisms indicate two possible fault plane solutions: (1) sinistral 075°/83° or dextral 165°/72° and (2) dextral 190°/71° or sinistral 100°/80° (Figure 2c). Although the strike variation between the fault plane solutions of the two average focal mechanisms is only 25°, we analyze these results separately and compare them with fitted fault planes and fault traces. The ENE-striking focal plane (075°/83°) of the first average focal mechanism has a similar orientation to the fitted planes of the collapsed events. It is also consistent in terms of both orientation and kinematics (sinistral strike-slip) with fault numbers 3a, 3b, and 4 of the fault model. Associating the second average focal mechanism to a causative structure in the fault model is more complicated and not unique. The N-striking dextral strike-slip focal plane (190°/71°) is, to a certain extent, similar to Fault 2 (Figure 1c); similarly oriented faults are recognized as responsible for earthquakes on the Reykjanes Peninsula (see, e.g., Keiding et al., 2009). Although the other focal plane (100°/80°) of this focal mechanism has the same orientation as structure number 8, this fault is likely to be dextral (Khodayar et al., unpublished report, Supporting Information, Table S3) and therefore does not fit with the focal mechanism solution. N-striking and ENE-striking planes could represent conjugate fault planes that have been observed elsewhere in Iceland (Khodayar et al., 2018). Based on fractures intercepted by the well and on computed focal mechanisms, the faults in the model are interpreted to be vertical.

### 2.3.3 Final revised fault model

Based on the analysis presented in the previous section, the final fault model consists of six vertical faults: one N-striking to NNE-striking structure (028°), Fault 2; one NNE-striking to NE-striking structure (034°), Fault 1; three ENE-striking structures (between 058° and 063°), Faults 3a, 3b, and 4; and one E-striking structure (strike of 100°), Fault 8 (Figure 3, left). ENE-oriented fault traces (Faults 3a, 3b, and 4 in Figure 1c) coincide with this interpreted surface expression of the boundary between the northern and the southern Riedel Shear Zones. They show a right-stepping en échelon arrangement similar to what has been observed elsewhere on the Reykjanes peninsula, typical of sinistral strike-slip kinematics (see Figure 1c). Uncertainties exist on the cross-cutting relationships (e.g., terminations or abutments) between the different fault sets (e.g., ENE-striking and NNE-striking to NE-striking), and interpretation in Khodayar et al. (2018) (section 2.1) did not match what was previously presented in a fault model scenario by Khodayar et al. (unpublished report, Supporting Information, Text S2). For example, in the former, ENE-striking faults are believed to cut NNE-striking to NE-striking volcanic fissure and normal faults, while in the latter, Fault 3b (ENE-striking) terminates on Fault 2 (NNE-striking to NE-striking). For this reason, cross-cutting relationships are not included in our fault model.

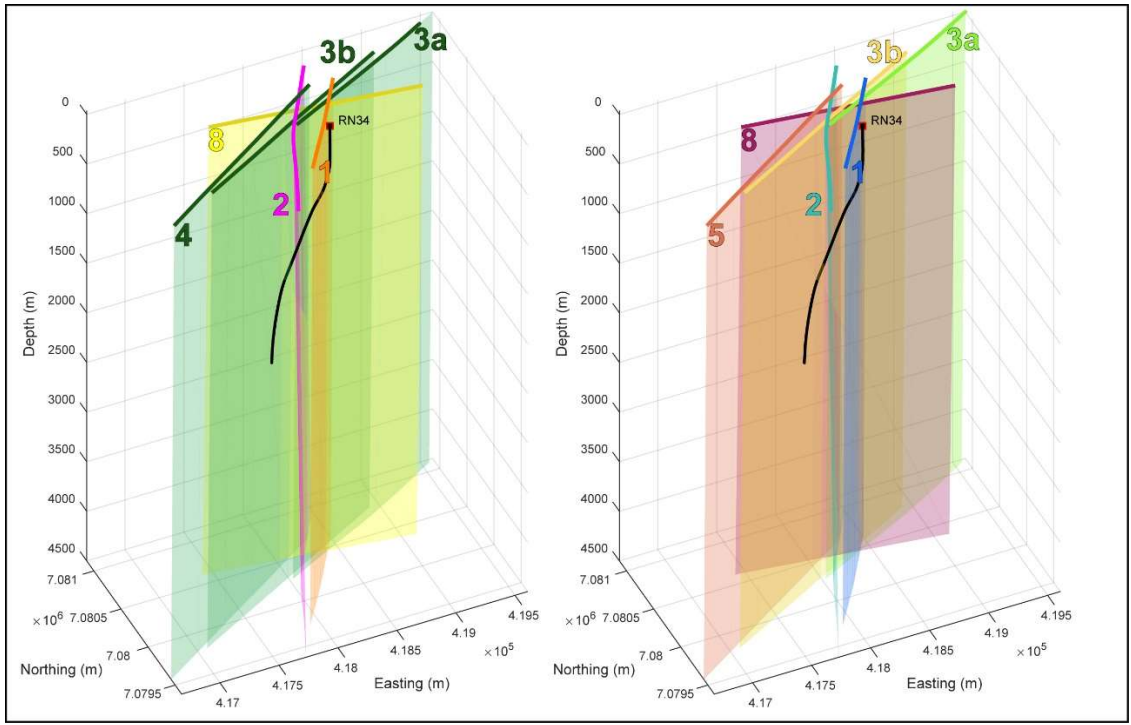

**Figure 3.** Fault model consisting of six faults. Left: Color scheme based on orientation and kinematics following unpublished work by Khodayar et al. (unpublished report, Supporting Information, Text S2). Right: Color scheme used for visualization of simulation results (see section 5), with different colors for each fault. Well location (square) and path (black line) are also shown.

### 2.4 Stress state

Focal mechanisms can also be used to infer the orientation of the stress state (see Figure 2d). The direction of the stress as derived from focal mechanism inversion is compared to strain rates from GPS data in Keiding et al. (2009) for different regions in Iceland and to the maximum horizontal stress derived from 15 orientation breakouts in RN-34 (Ziegler et al., 2016). The orientation for the western Reykjanes peninsula (extrapolated from Keiding et al., 2009, Figure 7) is approximately 220°/50° for the maximum principal stress axis $\sigma_1$, 064°/40° for the intermediate principal stress axis, and 323°/10° for the minimum stress axis $\sigma_3$. However, eastwards, the stress orientation changes toward the strike-slip regime of the SISZ. A model that advocated the permutation of two of the principal stress axes, $\sigma_1$ and $\sigma_2$, was initially used to explain oblique rifting, supported by the alternation of extension and strike-slip episodes and tectono-magmatic activity. However, the observed coexistence of normal, strike-slip, and oblique fault ruptures during both the swarm in 1972 and in 2013 (Khodayar et al., 2018; Klein et al., 1977) suggests a transtensional regime (intermediate case between normal and strike-slip regimes) in which differently oriented structures accommodate deformation.

For the 170 focal mechanisms derived in this study, we conducted a stress field inversion based on the method of Vavryčuk (2014); our resulting principal stress orientations are $\sigma_1$ (040° ± 5°)/23°, $\sigma_2$ (230± 10°)/66°, and $\sigma_3$ (140° ± 10°)/7°). The orientation of $\sigma_1$ is consistent with the strike of the maximum horizontal stress (034°) derived from breakouts in RN-34 (Ziegler et al., 2016). Note that the orientation of $\sigma_3$ is almost identical to that predicted by Keiding et al. (2009)

(Figure 2c, right), while $\sigma_1$ and $\sigma_2$ show the same trends but plunge on opposite quadrants. In addition, a series of stress inversion tests with varying friction coefficient shows that the most robust stress state is obtained for a friction coefficient of 0.4. This value is quite low compared to the 0.6–0.8 usually used (Byerlee, 1978). However, such a value has previously been observed when faults are partially filled with clays or low-friction minerals (Janecke & Evans, 1988; Kanji, 1974). In high temperatures, such as those present in the study region and relevant faults, decreased friction coefficients with temperature have been observed (e.g., by Di Toro et al., 2011).

Although information on stress orientation exists for the Reykjanes peninsula, direct measurements or estimates of stress intensity are lacking. Scenarios of four stress cases (corresponding to Andersonian's stress orientation plus a transtensional case) were accounted for by Peter-Borie et al. (2018) in their modeling of drilling effects and fracture initiation caused by stimulation of well RN-15/IDDP2 (circa 1.5 km SW of the study area). In their study, vertical stress intensity was estimated according to gravitational loading (134 MPa at a depth of 4560 m); and horizontal stress magnitudes were extrapolated from the stress state modelled by Batir et al. (2012) and Peter-Borie et al. (2018). The numerical model of stress was then compared to observations from well images suggesting that the strike-slip fault scenario and, to a lesser extent, the transtensional regime scenario predicted the breakouts more accurate.

## 3. The 29 March 2015 RN-34 fall-off test and cyclic well stimulation

This section describes available static measurements from well RN-34 as well as data from the 29 March 2015 testing and stimulation of the well. The operation consisted of two stages: a fall-off test followed by cyclic stimulation. For both stages, pressure and volume data are available, as is information from seismic monitoring.

### 3.1 Static and dynamic well data

RN-34 had been drilled to a depth of 2667 m on 27 March 2015 (Supporting Information, Text S3)[1]. Following rinsing of the well, a televiewer survey was conducted. Televiewer imaging indicated several possible feed points along the wellbore, with the main feed points most likely in the depth interval 2300–2600 m (Supporting Information, Text S1).

Testing of the well commenced on 29 March 2015 with a fall-off test followed by hydraulic stimulation. The fall-off test consisted of constant injection at a rate of 43 L/s from 07:15 to 09:50, followed by an abrupt shut-in. The pressure was monitored in the well at a depth of 1400 meters, that is, about 1000 m above the assumed leakage points from the well into the rock. Recording of data (shown in Figure 4b; see also Supporting Information, Text S4) started about 30 minutes before the shut-in and continued until 1.5 hours after shut-in (recording period 09:20–11:15). As can be seen from Figure 4b, the pressure was stable toward the end of the injection period and then decreased significantly after shut-in. The pressure drop between plateaus during and after injection was about 28 bar.

The following well stimulation was performed from 12:00 to 22:00 with a cyclic injection pattern with injection rates of 100 L/s applied for 1h, followed by rates of 20 L/s for periods of

---

[1] Drilling was completed at the final depth of 2695 m 2 April 2015 (Supporting Information, Text S3).

20–30 min (Supporting Information, Text S3). In the stimulations reported in Section 5, the durations of low injection rates were set to 30 min (see also the illustration in Figure 4c).

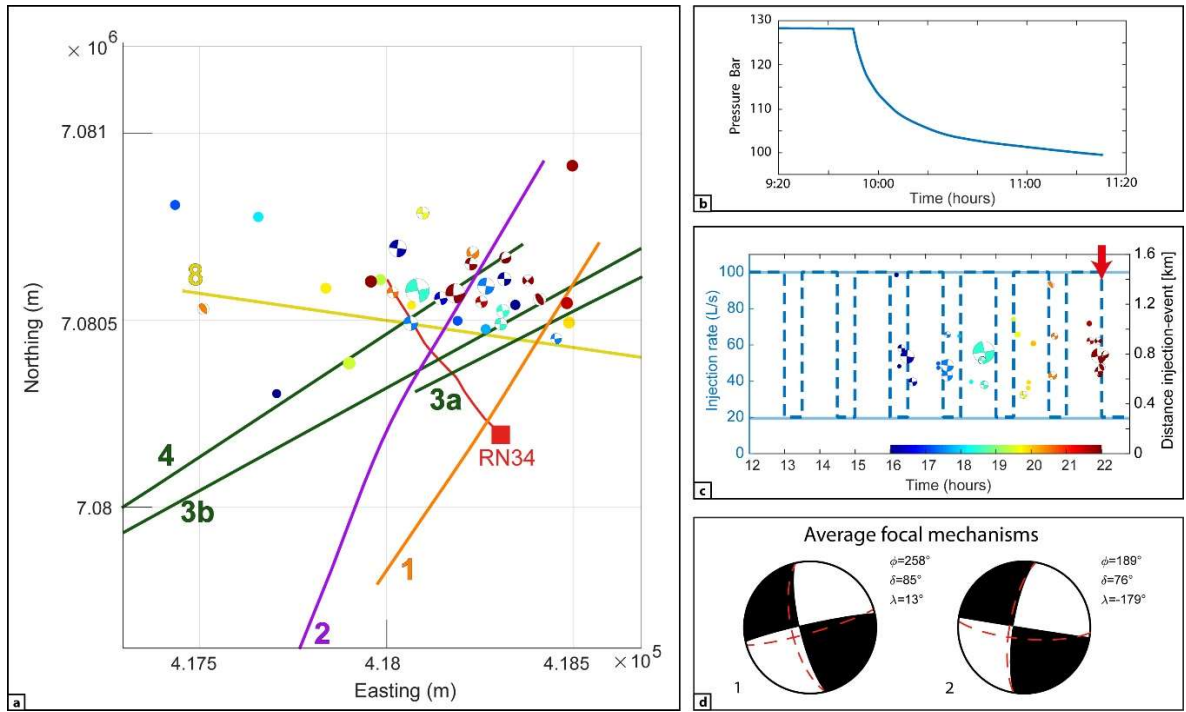

**Figure 4.** RN-34 injection test on 29 March 2015, seismic event occurrence, magnitude, and focal mechanisms. **(a)** Seismic event locations in map view, fault traces, and RN-34 well location. Colored dots represent seismic events; focal mechanisms are shown where determinable. The color scale applied to seismic events range in time from 16:00 (blue) to 22:00 (red); see subfigure (c). **(b)** Pressure evolution at the monitoring point during the fall-off test. **(c)** Injection rate and seismicity vs time, including the distance of a seismic event to the injection point. The applied injection rate was between 20 and 100 L/s, with periodic oscillations of 20–30 min intervals (Note: Detailed injection data not available). Dot and beach ball sizes are scaled by moment magnitude Mw 1.51 to 2.51. Red arrow corresponds to stimulation end. **(d)** Average of 19 focal mechanisms showing strike-slip kinematics. Red dashes show average focal planes from the three months of injection.

3.2 Induced seismic events

No seismicity was observed during the fall-off test. A sequence of 33 seismic events was observed in the period from 16:00 to 22:00, four hours after the stimulation characterized by the highest injection rate (100 L/s) started. The events are mainly located east of the injection point and north of Fault 8 (Figure 4a), consistent with observations during the three months of injection (Figure 1c). As the pressure and injection flux were not recorded at the well during the stimulation (Figure 4c), we could not link the triggered seismicity to any potential change in well pressure. However, the injection rates were higher than during the fall-off test. The proximity of the seismic events to the injection point (Figures 4a and 4c) and the fact that the events occurred during the stimulation phase suggest that they were induced by the fluid injection. The moment magnitudes vary from 1.5 to 2.5, which is within the range of the magnitudes estimated for the three months of seismic analysis ($M_w$ 0.8 to 3). However, the seismicity shows neither clear pattern (for example in terms of the spatial, temporal and magnitude distribution of events) nor correlation between locations and event occurrence time (Figure 4a). Moreover, the sequence of individual events does not show any clear increase in the distance to the injection point with time (Figure 4c) and does, hence, not follow a radial diffusion law (Shapiro et al., 2002).

The two principal focal mechanisms computed for 29 March 2015 (Figure 4d) show similar orientation and motion as the two principal focal mechanisms identified through the three-month period (20 May to 13 Aug 2015) of seismic analysis shown in section 2 (Figure 2b). With regard to the first principal focal mechanisms for both periods (29 March 2015 and the three-month period), one of the fault plane solutions is an ENE-striking plane with similar strike (258° vs. 75°) and dip (83° vs. 85°) but opposite dip-direction (north vs. south). The other fault plane solution is a N-striking plane with similar strike (165° vs. 168°) and dip-direction, but slightly different dip (85° vs. 72°). With regard to the second principal focal mechanisms for both periods, the N-striking plane solutions have a similar strike (190° vs. 189°), dip (76° vs. 71°), and dip-direction. These differences are not significant with respect to the relatively small number of events analyzed here. Thus, the reactivated structures during the stimulation on 29 March 2015 are likely the same as the structures reactivated during the latter three months of continuous injection.

To summarize, the sequence of 33 seismic events 29 March appears to be representative for the period from 20 May to 13 August in terms of spatial location and temporal occurrence, magnitude, focal mechanisms, and reactivated structures. This justifies our choice of the local fault model for this particular day as representative for a longer injection period. In addition, it also reinforces the hypothesis of vertical fractures used for the modeling.

## 4 Hydro-mechanical simulation tool

In this section we introduce a hydro-mechanical reservoir model to conceptualize the observations presented in sections 2 and 3 and to simulate fluid injection into a fault network as well as the mechanical response of the faults and the host rock to the fluid injection.

One main challenge in the numerical modeling of processes in faulted rocks is the large aspect ratio of faults and heterogeneity between faults and host rock. As the dominant physical processes in the faults are either different from those in the host rock or have substantially different characteristics, an upscaled representation that integrates host rock and faults into a continuous medium leads to models with poor accuracy. In particular, modeling of fracture reactivation and slip requires accounting for the deformation of the faults and the host rock and the coupling between them. How to incorporate this into a simulation model depends on whether the faults are resolved by the computational grid or not. Several studies have avoided resolving the fractures by applying subgrid-scale models to represent fracture-matrix interactions (Izadi & Elsworth, 2014; Norbeck et al., 2016; Rutqvist et al., 2015). Herein, we pursue a different approach, based on Discrete Fracture Matrix (DFM) principles (Berre et al., 2019), with the major faults explicitly represented in the computational grid. To avoid resolving the domain across the relatively thin faults , the faults are represented as lower-dimensional objects. The explicit representation gives transparent couplings of processes in host rock, fault network, and on the fault walls and, moreover, allows for high resolution of the sliding process. The effect of small-scale fractures, which are not explicitly represented, may be approximated by upscaling into matrix parameters.

While variants of DFM models have previously been applied to study shear stimulation of fault networks in geothermal reservoirs (Kolditz & Clauser, 1998; Sun et al., 2017; Ucar et al., 2017; Ucar et al., 2018a), this, to the best of our knowledge, is the first attempt at applying DFM models to simulation of coupled flow, mechanics, and fracture reactivation and slip for a case study from an actual geothermal reservoir.

The processes included in our simulation model are summarized as follows: The host rock is considered a poroelastic medium with a linear isotropic relation between stress and displacement. Fluid flow in the rock matrix and the fault network is modeled by Darcy's law. The deformation of the fault is modeled as a frictional contact problem between the fault walls: the fault can be open, in contact but sticking, or in contact and sliding. The latter is characterized by a jump in the tangential displacement of two opposing fault walls. This model is similar to previous models for poroelastic media with fractures modeled by contact mechanics considered recently (Berge et al., 2020; Gallyamov et al., 2018; Garipov et al., 2016; Garipov & Hui, 2019; Keilegavlen et al., 2019; Stefansson et al., 2020). Due to the limited data available to parameterize the model, the model applies only a constant friction coefficient and does not account for permeability enhancement due to shear dilation, although this could have been included (e.g., as by Stefansson et al., 2020). The full set of governing equations can be found in Supporting Information, Text S5.

The computational grid is constructed to conform to the explicitly represented faults. Faces on a fault surfaces are split, and lower-dimensional cells are inserted between the split faces (see Figure 5). The degrees of freedom in the simulation model are specified as illustrated in Figure 5: In the matrix grid, displacement and pressure are represented as cell-centered variables. Additional displacement degrees of freedom are placed on the faces on the fault surfaces. Finally, in the fault grid, fluid pressure and contact force (both normal and tangential) are represented by cell center values.

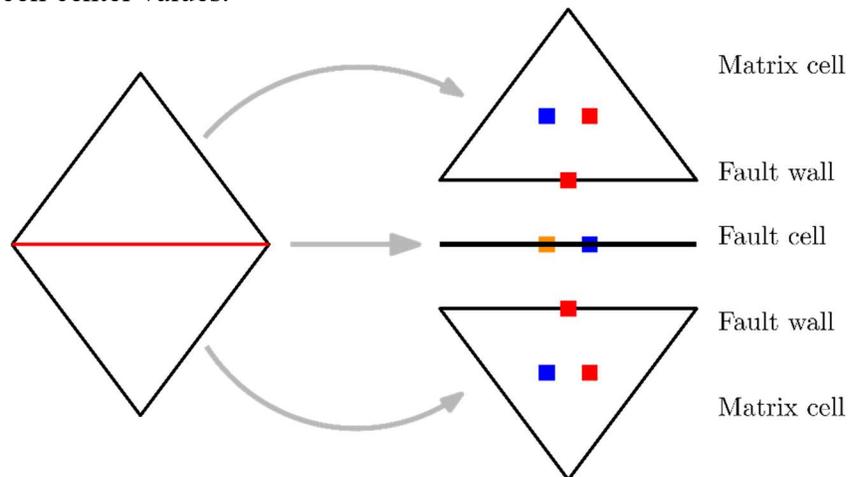

**Figure 5.** Illustration of conceptual model and degrees of freedom, shown in 2D for simplicity. Left: A DFM conceptual model, where the fault (red line) is represented as a lower-dimensional object. Right: Degrees of freedom: The deformation (red) is represented in matrix cells and on fault walls, fluid pressure (blue) in matrix and fault, and contact force (orange) in the fault.

The simulations are performed using the open-source simulator PorePy, described in Keilegavlen et al. (2019), and applying the following combination of discretization schemes: Poroelastic deformation of the matrix is discretized by the Multipoint Stress Approximation (Nordbotten, 2016; Nordbotten & Keilegavlen, 2020; Ucar et al., 2018b). Flow in matrix and faults is discretized by a Multipoint Flux Approximation (Aavatsmark, 2002), with the fluid flow between matrix and faults considered within the framework presented by Nordbotten et al. (2019). The simulation framework does not include a well-model but represents injection in a simplified manner by point sources in cells. The contact mechanics formulation for fault deformation uses a semi-smooth Newton approach, as described by Berge et al. (2020) and Hüeber et al. (2008).

## 5 Simulation results: Cyclic stimulation 29 March 2015 of RN-34

Our goal in this section is to present simulations of the stimulation event on 29 March 2015 and to compare the results with the analysis of seismic events as described in Section 5. To that end, a simulation model is constructed based on the geological information above.

### 5.1 Construction of simulation model

The geometry of the fault network is taken as described in section 2, and the full simulation domain is specified by a bounding box with a horizontal extent of 10x10 km. In the vertical direction, the simulation domain is set to 4 km. The faults are represented as being linear in the horizontal direction and assumed to be vertical and extending to the top and bottom boundaries of the domain (Figure 6). The simulation grid is created by first meshing the 2D horizontal domain with 1702 cells and then vertically extruding the grid with nine layers of non-uniform thickness, so that the zone near the injection has the highest grid resolution. The simulation grid for the matrix is illustrated in Figure 6.

The elastic moduli of the rock matrix are defined according to the seismic velocities of the rock, accounting for vertical variations of the rock properties; the assigned values are based on those reported by Bodvarsson et al. (1996) and given in the Supporting Information, Table S8. In accordance with section 2.4, the stress is assumed to be in a strike-slip regime, with the maximum and minimum principal stress directions both in the horizontal plane. Their magnitudes are taken as 1.5 times and 0.45 times the lithostatic stress, respectively, in accordance with Peter-Borie et al. (2018), with the maximum principal stress oriented in a NE-SW direction (see Figure 2d). These values are boundary conditions for the momentum conservation in the simulation model. The static friction coefficient on fracture surfaces was set to 0.4 in accordance with the analysis presented in section 2.4. The Biot coefficient was set to 0.8 and rock density to 3000 kg/m$^3$.

The parameters used in the flow model are fault and matrix permeability, matrix porosity, and the location of the feed points from the injection well into the formation. All of these parameters are both critical for the simulated formation response to the stimulation and highly uncertain.

As the feed points are assumed to be toward the bottom of the well (section 3.1) and associated with faults, the simulation model implements the feed point in the fault cell closest to the well at a depth of 2500 m. In practice, this places the feed point in Fault 4.

**Table 1**

*Hydraulic rock parameters for the three different cases.*

|  | **Case A** | **Case B** | **Case C** |
|---|---|---|---|
| Matrix permeability $K_M$ [m²] | 1e-12 | 2e-12 | 1e-11 |
| Hydraulic aperture $a$ [m] | 1e-2 | 1e-2 | 1e-2 |
| Tangential conductivity fault 1-4 | $a^3/12$ | $a^3/12$ | $a^3/12$ |
| Normal conductivity fault 1-4 | $a/6$ | $a/6$ | $a/6$ |
| Tangential conductivity fault 8 | $a^3/12$ | $a \cdot K_M$ | $1e-2 \cdot K_M \cdot a$ |
| Normal conductivity fault 8 | $a/6$ | $K_M/(a/2)$ | $1e-2 \cdot K_M/(a/2)$ |

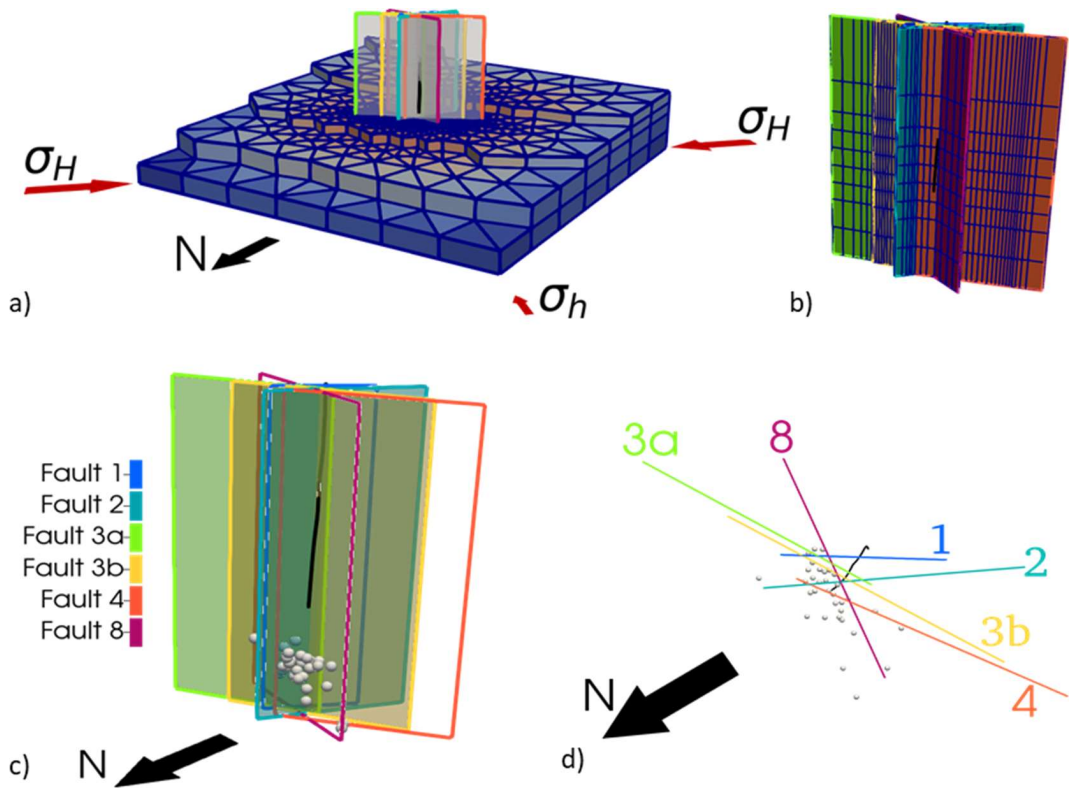

**Figure 6. (a)** The 3D simulation grid, split to expose the fractures. The coloring of the matrix cells shows the perturbation in pressure from the initial state for Case C. **(b)** The simulation grid in the fault network. **(c)** The fracture network with the color coding used for all visualization of simulation results, the well path (black), and the seismic observations (spheres). The orientation is chosen for optimal readability of the simulation results (Figures 7-9). **(d)** The fracture network, well path, and seismic observations seen from above.

The permeability values for the faults and matrix are reported in Table 1 The values are homogeneous in the matrix and for each of the faults and were held constant during the simulations. High permeabilities are assumed for Faults 1 to 4. The permeability of Fault 8 is considered as unknown prior to the simulation study, as its orientation relative to the regional stress field indicates it may have very low permeability and therefore practically sealing. Based on this, three scenarios were defined, with Fault 8 defined as permeable (Case A), as having permeability equal to that of the matrix (Case B), and as sealing (Case C). For each of these scenarios, the matrix permeabilities were tuned to reproduce the pressure drop measured in the leak-off test run on the morning of 29 March 2015. In this calibration, mechanical effects were ignored. While we acknowledge the simplicity of this approach, a richer parametrization and more elaborate calibration is not warranted due to the scarcity of data. Fluid properties are given in the Supplementary Information, Section 5. The flow simulation model is complemented by hydrostatic conditions at the lateral and upper boundaries and no-flow conditions at the bottom.

The hydro-mechanical simulation model is initialized by simulating a scenario with no injection until steady state is reached. As we have no information on the stress history of the reservoir, the stress boundary conditions in this initialization are taken as the current background stress field. Hence, the initial stress state of the faults in the simulation model may deviate from that of the reservoir, and fractures are critically stressed where slip has occurred. In the initialization, Faults 2 to 5 all undergo slip with a magnitude of order centimeters. Thereafter, the cyclic injection pattern described in section 3 is simulated with a time step of 15 minutes. In the discussion that follows, we mainly focus on the fault network; however, Figure 6 also shows the pressure perturbation in the matrix for Case C.

### 5.2 Simulation results

We first consider the initial slip tendency, defined as the ratio of tangential to normal forces on the fault surfaces (see Supporting Information, Text S5). As can be seen in Figure 7, the slip tendencies at the start of the stimulation for significant parts of Faults 2, 3a, and 3b attain the maximum possible value (equal to a fault friction coefficient of 0.4) and are thus critically stressed. Except for the tip of Fault 4, Faults 1, 4, and 8 have lower values. The slip tendencies along the faults undergo only minor changes during the stimulation. Hence, faults 1, 4, and 8 remain primarily uncritically stressed, and we report pressure profiles and tangential sliding for Faults 2, 3a, and 3b only. Of these, Fault 2 is connected to the injection point through the fault network without going through the potentially blocking Fault 8, while Faults 3a and 3b are favorably oriented with respect to the background stress field.

Figure 8 depicts pressure perturbations from the steady state at the end of the stimulation period (red arrows in Figure 4c). For Case A, the pressure perturbation is relatively high due to its lower matrix permeability, and pressure is diffused throughout the fracture network. In contrast, for Case C, the pressure perturbation is much lower and, to a large degree, localized in the part of Fault 2 that is on the same side of the sealing fault as the injection point. For the intermediate Case B, there is substantial pressure diffusion in the fault network due to the lack of a seal.

The slip along Faults 2, 3a, and 3b is shown in Figure 9. The slip profile is remarkably similar for the three cases, although the magnitudes of slip differ between them; the largest magnitudes were observed for Case A, which also has the most pronounced pressure diffusion in the fault network. The slip along Fault 2 for Case C is divided into a region close to the injection point and a region on the far (south) side of Fault 8. A similar division is not present in Cases A and B. Thus, for Case C it seems reasonable that the slip on the north side of Fault 8 is directly caused by fluid injection, while slip on the south side can be attributed to changes in the poroelastic stress in the surrounding rock matrix.

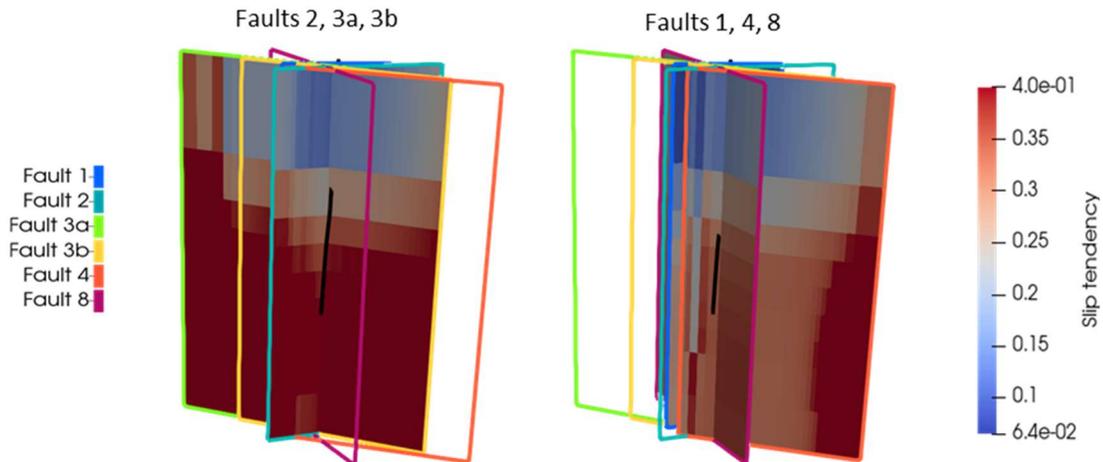

**Figure 7.** Slip tendency for Faults 2, 3a and 3b and Faults 1, 4 and 8 at the start of stimulation, computed for Case A.

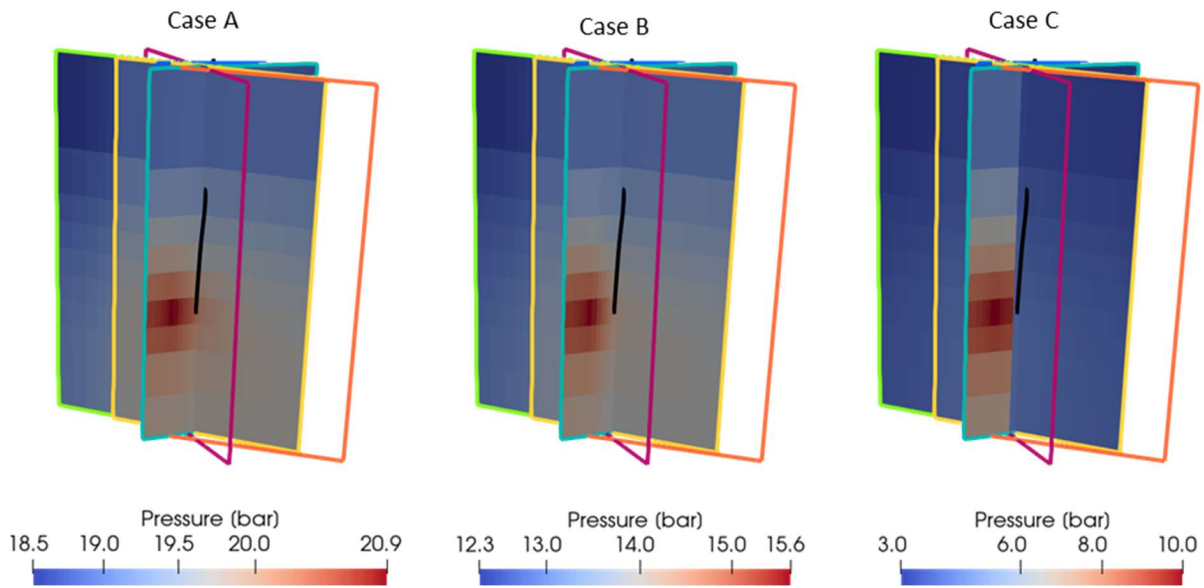

**Figure 8.** Difference in fault pressure between the end of the cyclic stimulation and the initial state for Case A, Case B and Case C. Note that the scale of the pressure color bar is different for each of the three cases.

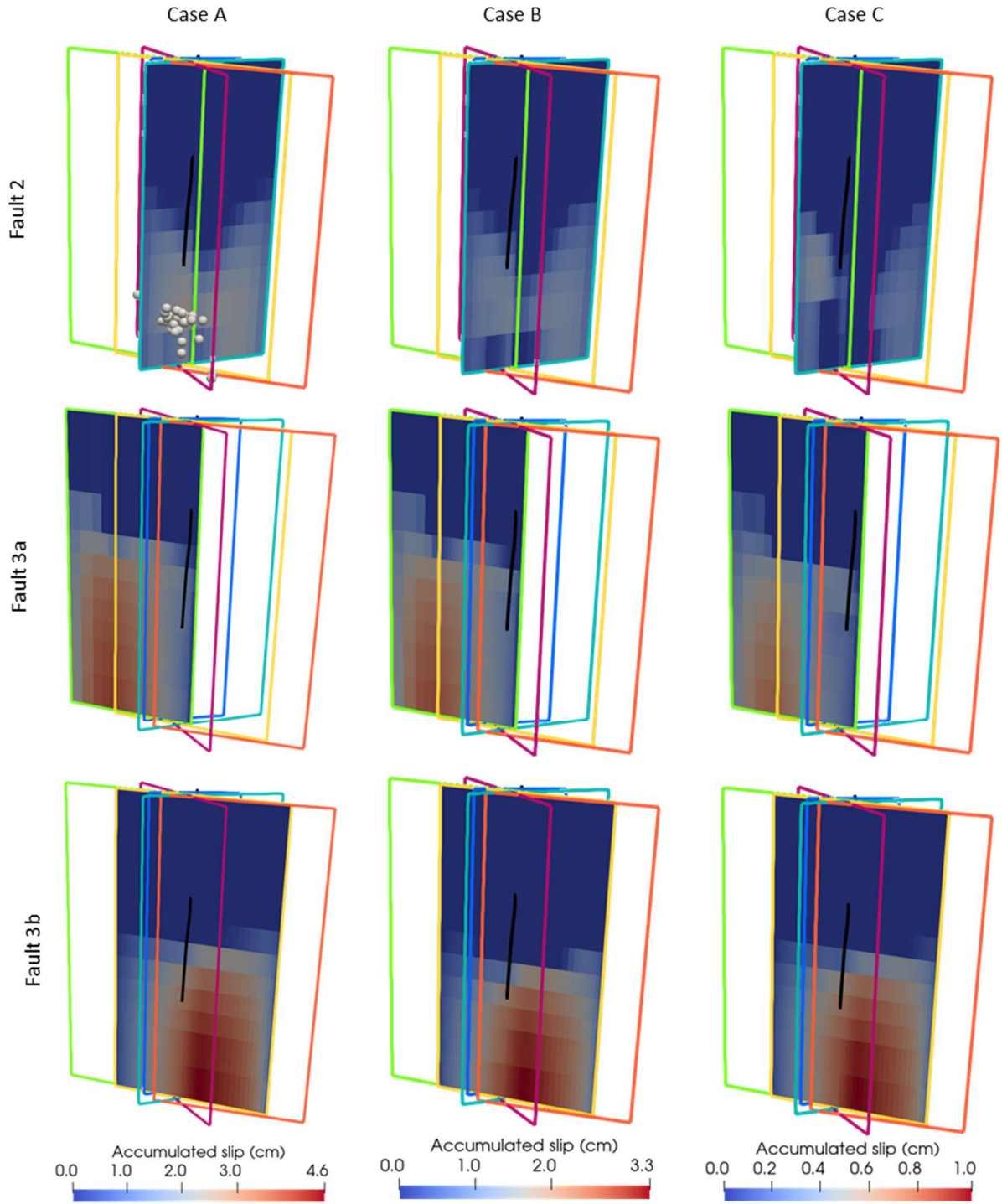

**Figure 9.** Accumulated slips at the end of the stimulation with Cases A, B, and C in the first, second, and third column respectively, and Faults 2, 3a, and 3b in the first, second and third row, respectively. The color bars (bottom) apply column-wise. Note that the scale of the color bar is different for each of the three cases. The upper left figure also shows the observed seismic events, as discussed in section 3.2. For the legend for fault numbers and orientation of north, see Figure 6.

## 6 Discussion

During the cyclic stimulation of RN34 on 29 March 2015, 33 seismic events were observed. The events were mainly located east of the injection point, north of Fault 8, and below the bottom of the well. For the given simulation model, the main active faults are Faults 2, 3a, and 3b, showcasing how modeling is able to discriminate slip on close-by and similarly oriented faults (e.g., 3a, 3b, and 4), where seismicity fails because location uncertainties are larger than stepover width. Regions of main simulated slip along these faults are consistent with observed seismicity during stimulation of RN-34. At the same time, the faults also have simulated slip in regions farther from the seismic cloud. However, simulated slip in regions where little seismic activity is observed is not a clear proof of model deficiency, as slip in this region could be aseismic. Evidence of aseismic slip is, e.g., observed in the Brawley (Wei et al., 2015) or Soultz-Sous-Forêts (Cornet et al., 1997) geothermal fields; and in decameter-scale experiments (De Barros et al., 2019; Duboeuf et al., 2017; Guglielmi et al., 2015). Seismic analysis (section 2.3) of the data from the latter three-month injection period, 20 May to 13 August 2015, identifies the ENE-striking faults (Faults 3a, 3b, and 4) as most likely being the active faults. Simulation results show that while Faults 3a and 3b are active, Fault 4, located in the shadow of the nearby active faults, is mainly inactive. However, Fault 4 has some slip and a high slip tendency, in particular near the tip. Shadow effects can also be seen on Fault 2, which has less slip in the region close to where it is intersected by Faults 3a and 3b. Hence, when the location of seismicity is uncertain in regions of nearby and similarly oriented faults, modeling allows us to discriminate on which fault slip occurs.

The discussion above shows that the simulation model can contribute to the understanding of coupled hydromechanical processes interacting with deformation along preexisting faults. At the same time, the current model has severe limitations. While a friction model incorporating stable and unstable slip could be introduced in the simulation model to investigate this issue further, the available data is insufficient to parametrize such a refined model. While the local stress field before stimulation has strong influence on the extent to which faults are reactivated during the stimulation, the initialization process for the simulations is based on strong approximations, which are in turn based on fixed stress boundary conditions and a simple friction model. These aspects, along with the high uncertainty in model parameters, indicate significant model error.

Among these uncertainties, fault network geometry plays an important role as it can strongly affect connectivity. For example, our model representation of faults as linear in the horizontal direction has Faults 2 and 3a as crossing; this may not be the case based on fault traces mapped at surface. Furthermore, a cross-cutting relationship between different fault sets at depth largely affects connectivity. However, modeling allows testing of different geometries, comparing the outcomes, and linking them with observations. As the seismic cloud during stimulation is located mainly to the north of Fault 8, on the same side of the fault as the fluid injection, a simple scenario-based study was designed to investigate the effect of varying the relative permeability of Fault 8 compared to the other faults and the matrix. For Case A, Fault 8 had a high permeability equal to that of the other faults; in Case B, it had a permeability equal to that of the matrix; and in Case C, it had a permeability significantly lower than the matrix. There is significant difference in maximum accumulated slip along the active faults in the different cases, with longest slip in Case A and shortest in Case C. In Cases A and B, the regions of the faults that have slipped are similar and have similar slip profiles. The reason is likely that the faults are already critically stressed when injection starts, so that only small change in fluid pressure or poroelastic stress state will induce additional slip, which will redistribute poroelastic stress and further affect slip tendencies. For Case C, the effect of Fault 8 as a barrier to flow and the higher matrix permeability relative to Cases A and B can clearly be seen on Fault 2. The higher matrix permeability results in pressure migrating more easily into the matrix and lower pressures in all faults, including Fault 2 as compared in Cases A and B. At the same time, considering the large difference in results when changing the permeability of structures in the formation, slip distances along the active faults are clearly sensitive to the permeabilities of the faults and the matrix. Due to limitations in available data, the permeability of all faults (except Fault 8 in Cases B and C) was set as equal. Given that the NNE-striking to NE-striking structures (including Fault 1) display an aperture in the televiewer and that Fault 1 was deduced as open from the outcrop study (Khodayar et al., unpublished report, Supporting Information, Text S2), an alternative scenario could differentiate these fault sets from those that are ENE-striking (Faults 3a, 3b, and 4). This scenario is further supported by considering the normal load on the two sets of structures. Furthermore, the model is set up considering explicit representation of only six planar faults, while the actual fault geometry of the formation is richer, with larger-scale structures that are not resolved with the current model. Finally, the choice of a static friction and ignoring dilation of fractures with slip is also a simplification of the real situation.

## 7 Conclusion

Combining analysis of seismicity observed during well stimulation with simulation of injection-induced reservoir dynamics has the potential to improve our understanding of injection-induced fault reactivation as well as interpretations of data. Considering a case study from Reykjanes, Iceland, we have presented a workflow where we first used new analysis of seismic data to establish a revised fault model before this model was used in simulation experiments. Reciprocally, the simulation results show how modeling can be used as a tool to improve interpretations from seismic analysis, e.g., in discriminating slip along close-by and similarly oriented faults. The test cases investigated also show how sensitive fault slip is to the initial stress state as well as the permeability of the faults and their surrounding formation.

Limitations of the current work are related to uncertainty in geological characterization and seismic analysis and model error. While the model framework allows for the introduction of more complex physics, this would lead to over-parameterization as the data are insufficient for identification of the additional parameters. For the seismic analysis, downhole monitoring instrumentation would allow for more precise event locations that could be used to inform the fault model. Furthermore, additional data from pressure transient testing with downhole pressure measurements would improve calibration of permeabilities for the different structural components of the model.

## Acknowledgements and Data

HS Orka is acknowledged for allowing information from their reports produced by ÍSOR to be published in the Supporting Information (Texts S1-4 and Tables S1-7) to this paper.

Source code and run scripts for the simulations in this paper are openly available under Version 3 of the GPL license (Keilegavlen & Stefansson 2020).

Seismic data will be made available upon publication of the manuscript.